\documentclass[reprint,
superscriptaddress,
amsmath,amssymb,aps,showkeys,showpacs,
twoside,final,secnumarabic,
nofootinbib]{revtex4-2}

\usepackage[paperwidth=205mm,paperheight=290mm,top=17mm,bottom=25mm,
inner=17mm,outer=17mm,
twoside]{geometry}

\usepackage{cmap} 
\usepackage[T1,T2A]{fontenc}
\usepackage[utf8]{inputenc}
\usepackage[russian,english]{babel}
\usepackage{color}
\usepackage{graphicx}
\usepackage{dcolumn}
\usepackage{bm} 
\usepackage[unicode=true,colorlinks=true,linkcolor=magenta, urlcolor=blue, citecolor = blue,breaklinks]{hyperref}
\usepackage{multirow}
\usepackage{url}
\usepackage{breakurl}
\DeclareGraphicsExtensions{.eps}

\newcount\issue
\newcount\Vol
\newcount\numb
\usepackage{fancyhdr} 
\def\Vol{\textbf{78}}
\def\numb{x}
\begin{document}

\title{
Hyperon Physics at BESIII} 

\def\addressa{School of Physical Science and Technology, Lanzhou University, Lanzhou 730000, People’s Republic of China}
\def\addressb{Lanzhou Center for Theoretical Physics, Frontiers Science Center for Rare Isotopes, Key Laboratory of Theoretical Physics of Gansu Province and Key Laboratory for Quantum Theory and Applications of MoE, Lanzhou University, Lanzhou 730000, People’s Republic of China}

\author{\firstname{Xiongfei}~\surname{Wang} (on the hehalf of BESIII Collaboration)}
\email[E-mail: ]{wangxiongfei@lzu.edu.cn}
\affiliation{\addressa}\affiliation{\addressb}

\received{xx.xx.2024}
\revised{xx.xx.2024}
\accepted{xx.xx.2024}

\begin{abstract}
The BESIII detector on the BEPCII collider collected the world's largest dataset at the peaks of $J/\psi$, $\psi(3686)$ and $\psi(3770)$.
The use of polarization and entanglement states in multidimensional angular distribution analysis can provide new probes to the production and decay characteristics of hyperon anti hyperon pairs. 
In a recent series of studies, significant transverse polarization in hyperon decay has been observed in $J/\psi$, $\psi(3686)$ decaying into the $\Lambda\bar\Lambda$, $\Sigma^+\bar\Sigma^{-}$, $\Xi^0\bar\Xi^0$ and $\Xi^-\bar\Xi^{+}$ final states. The weak decay parameters of hyperons and antihyperons are also independently determined for the first time. 
The most accurate testing for direct $CP$ violation has been achieved.

\end{abstract}

\par
\keywords{BESIII experiment, Hyperon pair production, Hyperon spin polarization, $CP$ violation   \\[5pt]}

\maketitle
\thispagestyle{fancy}

\section{Introduction}\label{intro}
Standard Model (SM) of particle physics  has achieved great success in describing elementary particles and their interactions, but there are still some unresolved issues.
One of key challenge is to figure out why there is much more matter than antimatter in the Universe.
The violation of the joint charge-conjugation and parity ($CP$) symmetry is one of three essential conditions in Sakharov’s theory for understanding the matter-antimatter asymmetry in the Universe~\cite{Sakharov:1967dj}. 
Despite the established presence of $CP$ violation ($CPV$) in the decays of kaon, charm and beauty meson decays~\cite{Workman:2022ynf}, the SM of particle physics, as described by the Cabibbo–Kobayashi–Maskawa formalism, is insufficient in completely explaining the preponderance of matter over antimatter in the Universe~\cite{Peskin:2002mm}.
As a result, all observated results did not exceed the SM predictions with a small $CPV$, which cannot explain the asymmetry between large matter and antimatter in the Universe~\cite{Cohen:1993nk}

Two-body hyperon pair production and subsequent weak decays 
in Charmonium states $J/\psi$, $\psi(3686)$, $\psi(3770)$, and $e^+e^-$ collisions provide a rich laboratory and ideal probes in the study of fundamental symmetry properties in particle physics~\cite{BESIII:2019dve, BESIII:2019cuv, BESIII:2020ktn, BESIII:2021aer, BESIII:2021gca, BESIII:2021ccp, BESIII:2022mfx, BESIII:2022kzc, BESIII:2023rse}. Among them, the nonleptonic decays of spin-$\frac{1}{2}$ hyperons are suitable for $CPV$ studies. In such decays, the decay asymmetry parameters $\alpha$, $\beta$ and $\gamma$ are defined in terms of the S-wave (parity violating) and P-wave (parity conserving) amplitude contributions, and only two of them are independent~\cite{Lee:1957qs, Liu:2023xhg}.

In this proceeding, we present the recent progresses on the experimental studies of hyperon spin polarization and $CPV$ in $\Lambda$, $\Sigma^+$, $\Xi^0$ and $\Xi^-$ decays at BESIII.  More hyperon spin polarizations are observed and the test of $CPV$ is reported with a precision of $10^{-3}$. 

\section{\label{sec:level1} Recent results}
\subsection{Hyperon physics in $\Lambda$ decay}
\subsubsection{$J/\psi\to\Lambda\bar\Lambda$}
\begin{figure}[!htbp]
\begin{center}
\includegraphics[width=0.23\textwidth]{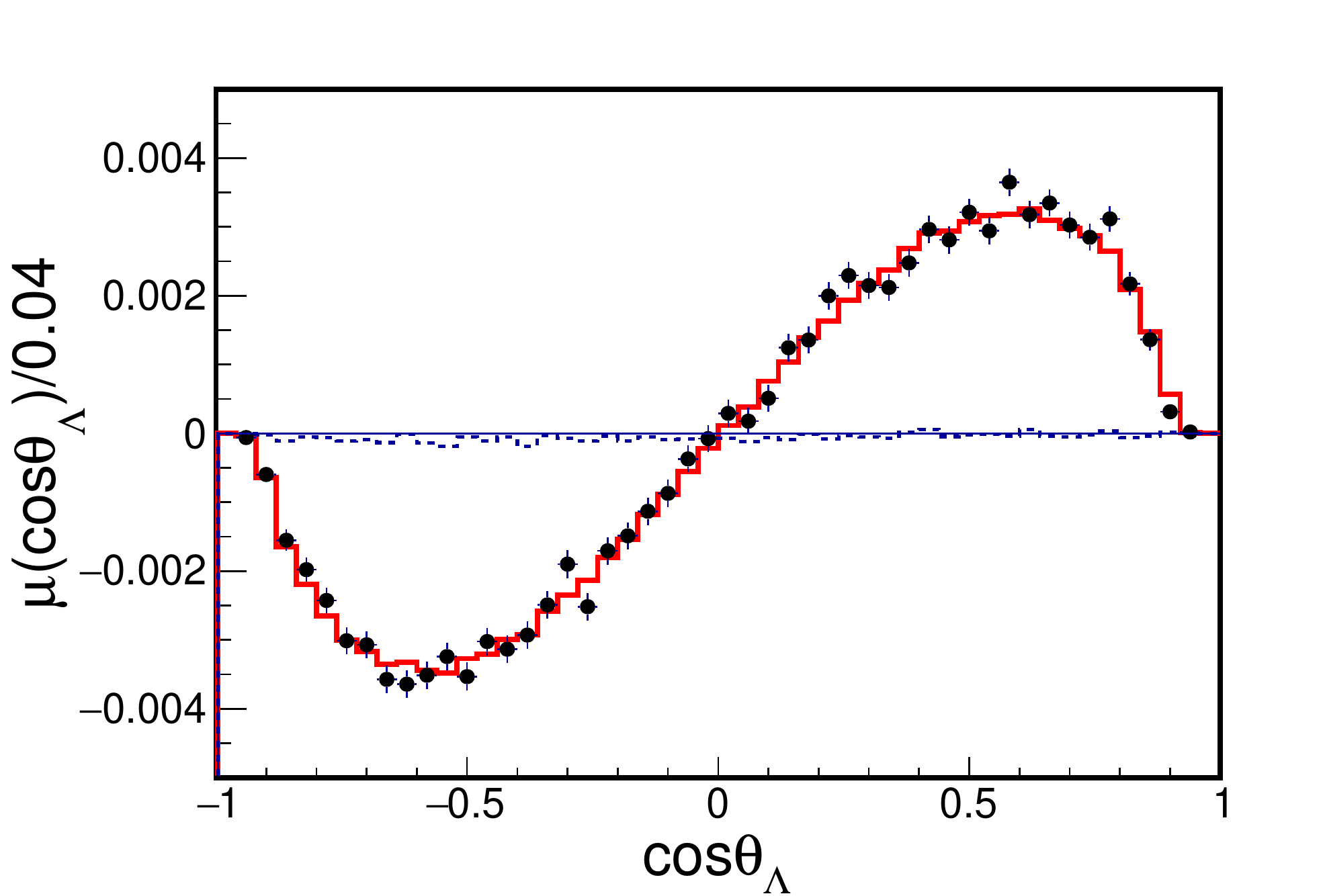}
\includegraphics[width=0.23\textwidth]{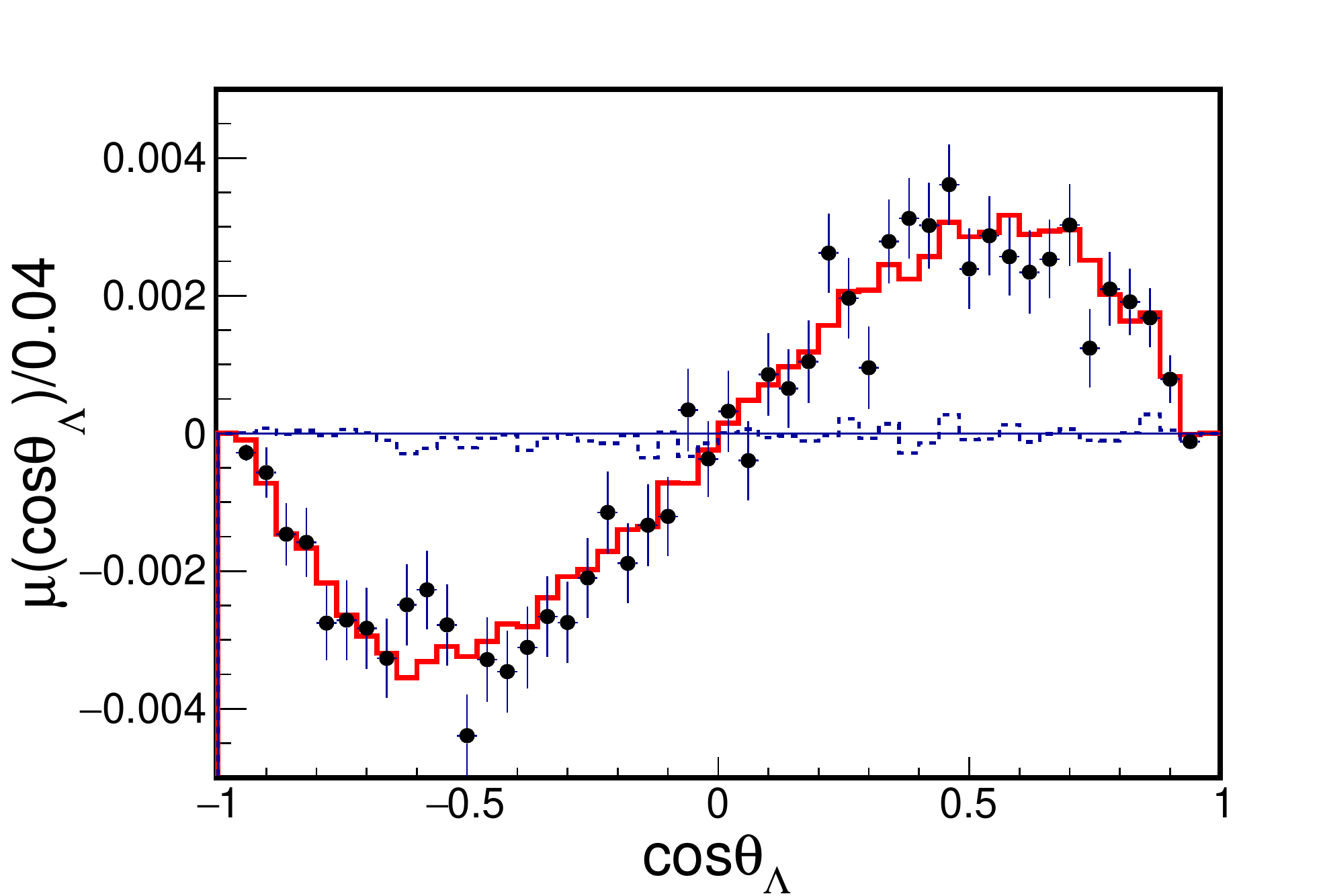}
\end{center}
\caption{\label{scatter_plot::llb}The moments $\mu(\cos\theta_{\Lambda})$ as a function of $\cos\theta_{\Lambda}$ in $J/\psi\to\Lambda\bar\Lambda$ for $p\pi^{-}\bar{p}\pi^{+}$ (Left) and $p\pi^{-}\bar{n}\pi^{+}$ (Right) final states.}
\end{figure}

In 2019, BESIII experiment proposed a novel way to probe the $CPV$ using the quantum entangled $\Lambda\bar\Lambda$ pairs in $J/\psi$ decay with a five dimension angular distribution analysis~\cite{BESIII:2018cnd}. Clear $\Lambda$ hyperon spin polarization has been observed for the first time as shown in Fig.~\ref{scatter_plot::llb}.
The nonzero value of $\Delta\Phi$ allows for a direct and si­multaneous measurement of the decay asymmetry pa­rameters.
The decay parameter $\alpha_{-}$ for $\Lambda\to p\pi^-$ is measured to be 17\% higher than the PDG value with the 5$\sigma$ difference.
The decay parameter $\bar\alpha_{0}$ for $\bar\Lambda\to\bar{n}\pi^+$ is measured for the first time. The ratio $\bar\alpha_{0}/\bar\alpha_{+}$ is 3$\sigma$ smaller than unity, and indicates an isospin $\Delta I =\frac{3}{2}$ contribution to the final state~\cite{Overseth:1969bxc, Olsen:1970vb}.
Other decay parameters are measured with high precision, and the most accurate test of $CPV$ is obtained in $\Lambda$ hyperon decay by then.

With the successful completion of 10 billion $J/\psi$ events collection, the spin polarization and the test of $CPV$ have been updated 
with most precise measurement~\cite{BESIII:2022qax}. Results are consistent with the previous measurements of BESIII experiment~\cite{BESIII:2018cnd}, and with higher precision. The $CPV$ is measured with an order of $10^{-3}$, which still has a long road to reach the SM prediction.
Table~\ref{table:sum_decay:llb} summarizes the measured results. 

\subsubsection{$\psi(3686)\to\Lambda\bar\Lambda$}
The $\Lambda$ psin polarization in the $e^+e^-\to\Lambda\bar\Lambda$ reaction was also studied
with data samples collected at seven energy points at
$\sqrt{s}$ = 3.68-3.71 GeV, corresponding to an integrated luminosity of 333 $pb^{-1}$
~\cite{BESIII:2023euh}.
The relative phase and the modulus of the ratio of the EM-psionic form factors and the angular distribution parameter are determined with the assumption of $CP$ conservation.  The relative phase is found to be different from zero with a significance of 2.6$\sigma$ including the systematic uncertainty. Table~\ref{table:sum_decay:llb} summarizes the measured results. These results provide more information for understanding the production mechanism of $\Lambda\bar\Lambda$ in $\psi(3686)$ and the internal structure of $\Lambda$ hyperon.

\subsubsection{$\psi(3770)\to\Lambda\bar\Lambda$}
Using a data sample of $\psi(3770)$ events corresponding to an integrated luminosity of 2.9 fb$^{-1}$,  the measurement of $\Lambda$ spin polarization in $e^+e^-\to\Lambda\bar\Lambda$ at $\sqrt{s} = 3.773$ GeV also has been attempted~\cite{BESIII:2021cvv}. 
The relative phase and the modulus of the ratio of the EM-psionic form factors and the angular distribution parameter are determined with the assumption of $CP$ conservation as listed in Table~\ref{table:sum_decay:llb}.
The measured phase is found to be different from zero with a significance of 2$\sigma$ including the systematic
uncertainties. 
The results also provide more information for understanding the $\Lambda$ pair production mechanism in $\psi(3770)$ decay and the internal structure of $\Lambda$ hyperon.

\subsubsection{$e^+e^-\to\Lambda\bar\Lambda$}
With an integrated luminosity of 66 $pb^{-1}$ data sample taken at $\sqrt{s} =$ 2.396 GeV, 
the $\Lambda\bar\Lambda$ pair production and subsequent $\Lambda$ decay near $\Lambda\bar\Lambda$ mass threshold were studied~\cite{BESIII:2019nep}. 
A nonzero spin polarization is observed, which confirms the previous observation of $\Lambda$ spin polarization in other energy points. 
The first complete determination of $\Lambda$ EMFF is reported.
In addition, the Born cross section and effective form factor for $e^+e^-\to\Lambda\bar\Lambda$ at $\sqrt{s} =2.396$ GeV are measured. Table~\ref{table:sum_decay:llb} summarizes the measured results. 
More information for understanding $\Lambda\bar\Lambda$ pair production and the $\Lambda$ internal structure near threshold are provided.
\begin{table}[htbp]
        \centering
        \caption{Experiment parameters for angular distribution ($\alpha$), spin polarization ($\Delta\Phi$), decay parameters ($\alpha_{\mp}$, $\alpha_{0}$), $CP$ observable ($A_{CP}$) and EMFF ratio ($R^{(\Psi)}$) in $J/\psi$, $\psi(3686)$, $\psi(3770)$ decays and in $e^+e^-$ annihilation.}
  \scalebox{0.56}{
        \begin{tabular}{c|c|c|c|c}
        \hline
        \hline
        Para.          &$J/\psi\to\Lambda\bar\Lambda$~\cite{BESIII:2018cnd, BESIII:2022qax}    &$\psi(3686)\to\Lambda\bar\Lambda$~\cite{BESIII:2023euh}    &$\psi(3770)\to\Lambda\bar\Lambda$~\cite{BESIII:2021cvv}&$e^+e^-\to\Lambda\bar\Lambda$~\cite{BESIII:2019nep}  \\ 
        \hline
        $\alpha$                              &$0.4748 \pm 0.0022 \pm 0.0031$   &$0.69 \pm 0.07 \pm 0.02$   &$0.85^{+1.2}_{-2.0} \pm 0.02$  &$0.12 \pm 0.14 \pm 0.02$\\
        $\Delta\Phi$ ($^\circ$)          &$43.09 \pm 0.24 \pm 0.38$           &$23^{+8.8}_{-8.0} \pm 1.6$  &$71^{+66}_{-46} \pm 5$           &$37 \pm 12 \pm 6$   \\
        $\alpha_{-}$                         &$0.7519 \pm 0.0042 \pm 0.0066$   &0.7539 (fixed)                      &0.754 (fixed) &--\\
        $\alpha_{+}$                         &$-0.7559 \pm 0.0036 \pm 0.0024$ &-0.7539 (fixed)                     &-0.754 (fixed) &--\\
        $\bar\alpha_{0}$                   &$-0.692 \pm 0.016 \pm 0.006$       &--&--&--\\
        $A_{CP}$ ($\times 10^{-3}$)   &$-2.5 \pm 4.6 \pm 1.2$                  &0&0&0\\
        $\bar\alpha_{0}/\alpha_{+}$   &$0.915 \pm 0.022 \pm 0.008$&--&--&--\\
         $R^{(\Psi)}$                         &$0.82823 \pm 0.0024 \pm 0.0033$ &$0.71 \pm 0.10 \pm 0.03$ &$0.48^{+0.21}_{-0.35} \pm 0.04$  &$0.96 \pm 0.14 \pm 0.02$\\
        \hline
        \hline
        \end{tabular}}
        \label{table:sum_decay:llb}
\end{table}
\subsection{Hyperon physics in $\Sigma^+$ decay}
\subsubsection{$J/\psi$, $\psi(3686)\to\Sigma^+\bar\Sigma^-\to p\bar{p}\pi^0\pi^0$}
Using 1.3 billion $J/\psi$ and 448 million $\psi(3686)$ events taken by BESIII detector, 
the $\Sigma^{+}$ spin polarization was reported for the first time with a five dimensional angular analysis of the process of
$J/\psi$, $\psi(3686)\to\Sigma^+\bar\Sigma^-\to p\bar{p}\pi^0\pi^0$~\cite{BESIII:2020fqg}.
Clear $\Sigma^+$ hyperon spin polarization has been observed as shown in Fig.~\ref{scatter_plot::ssb}.
The $\bar\alpha_{0}$ for $\bar\Sigma^-\to\bar{p}\pi^0$ is measured for the first time, while the $\alpha_{0}$ for $\Sigma^+\to p\pi^0$
is measured to be consistent with the previous result, and with significantly improved precision.
Table~\ref{table:ssb} summarizes the measured results. 
The $CPV$ observable is extracted based on the measured decay parameters in $\Sigma^+$ hyperon decay, and still favors the $CP$ conservation under the current statistics.

\subsubsection{$J/\psi\to\Sigma^+\bar\Sigma^-\to p\pi^0\bar{n}\pi^{-} + c.c.$}
With the same strategy as the $\Sigma^+\to p\pi^0$ mode~\cite{BESIII:2020fqg},  a five dimensional angular analysis of the process $J/\psi\to\Sigma^+\bar\Sigma^-\to p\pi^0\bar{n}\pi^{-} +c.c.$ was also implemented based on 10 billion data sample~\cite{BESIII:2023sgt}.
The decay parameter for $\bar\Sigma^{-}\to\bar{n}\pi^-$ is measured for the first time and other parameters are measured to be consistent with the previous measurement, and with improved precision. The nonzero value of $\Delta\Phi$ is observed with larger significance as shown in Fig.~\ref{scatter_plot::ssb}, and confirms that the observation of $\Sigma^+$ transverse polarization in $\Sigma^+\to p\pi^0$ mode. 
Table~\ref{table:ssb} summarizes the measured results. 
The study of $CPV$ for $\Sigma^+$ hyperon to the neutron decay is performed and result is consistent with $CP$ conservation.
\begin{figure}[!htbp]
\begin{center}
\includegraphics[width=0.23\textwidth]{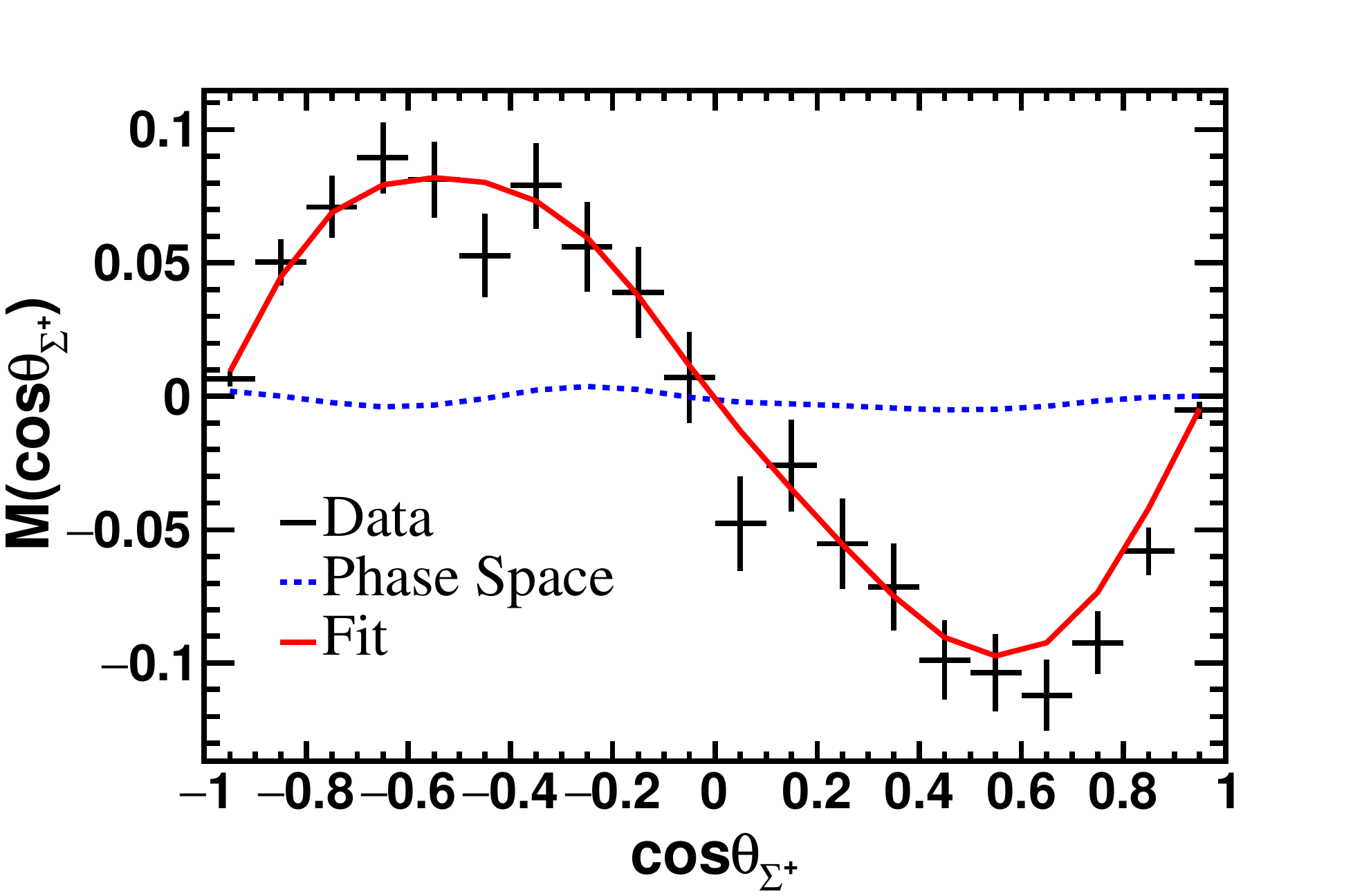}
\includegraphics[width=0.23\textwidth]{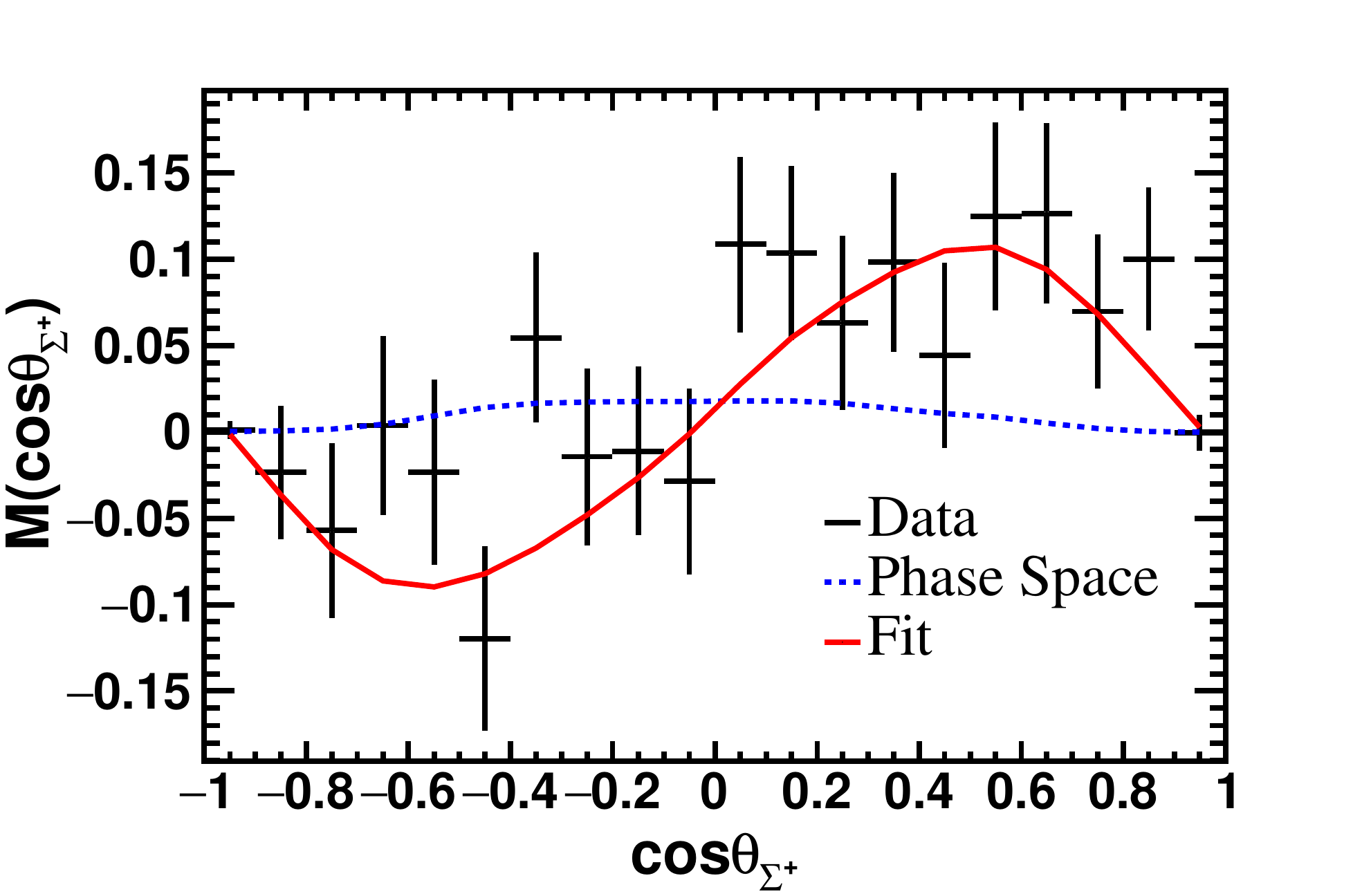}
\includegraphics[width=0.25\textwidth]{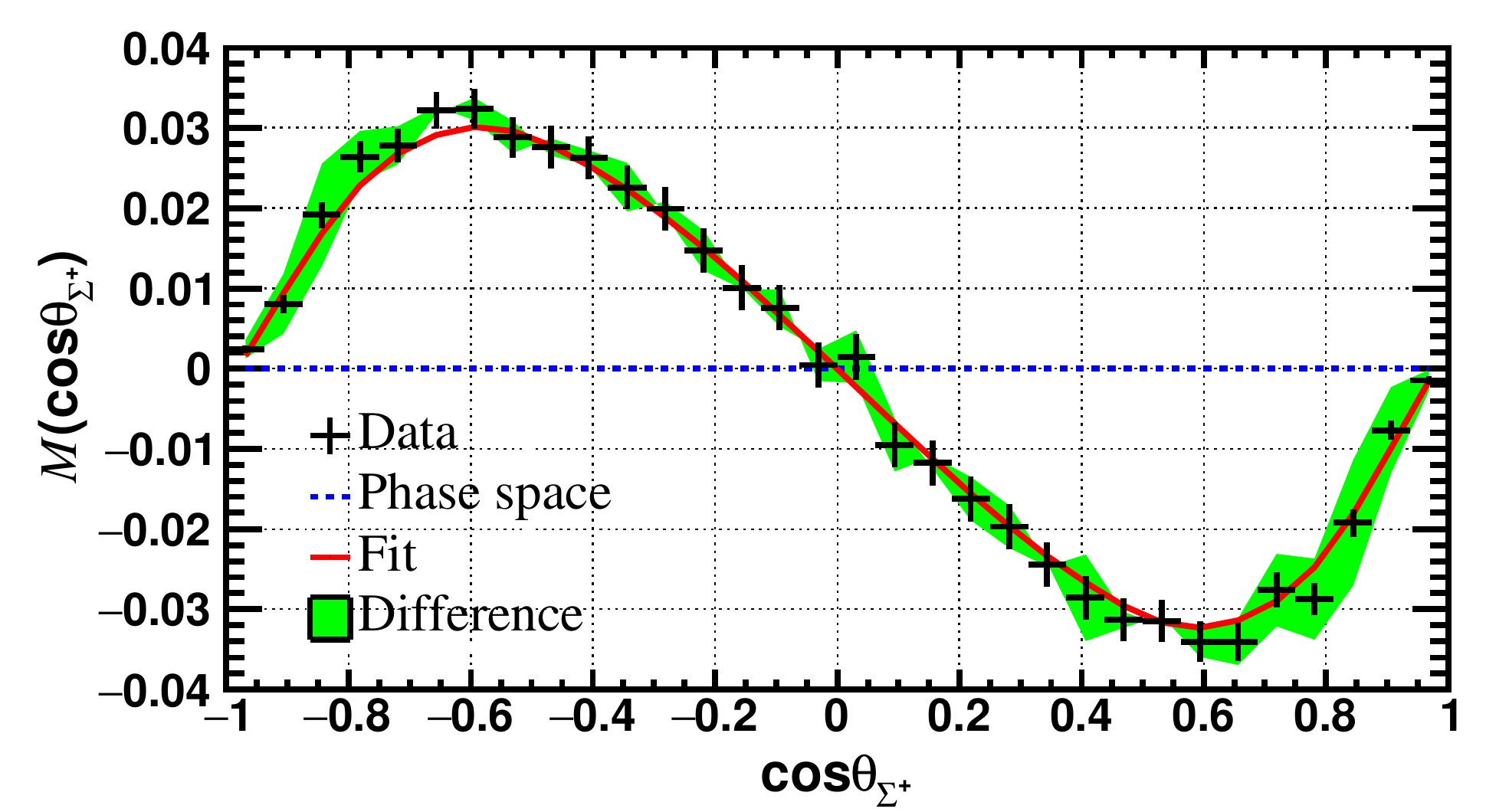}
\end{center}
\caption{\label{scatter_plot::ssb}The moments $\mu(\cos\theta_{\Sigma^+})$ as a function of $\cos\theta_{\Sigma^+}$ in $J/\psi\to\Sigma^+\bar\Sigma^-$ for $p\pi^{0}\bar{p}\pi^{0}$ (Top Left) and $p\pi^{0}\bar{n}\pi^{-}$ (Bottom) final states, and $\psi(3686)\to\Sigma^+\bar\Sigma^-$ for  $p\pi^{0}\bar{p}\pi^{0}$ (Top Right). }
\label{scatter_plot::ssb}
\end{figure}
\begin{table}[htbp]
        \centering
        \caption{Experiment parameters for angular distribution ($\alpha$), spin polarization ($\Delta\Phi$), decay parameters ($\alpha_{\mp}$, $\alpha_{0}$($\bar\alpha_{0}$)), $CP$ observable ($A_{CP}$) and EMFF ratio ($R^{(\Psi)}$) in $J/\psi$ and $\psi(3686)$ decays, and in $e^+e^-$ annihilation.}
  \scalebox{0.63}{
        \begin{tabular}{c|c|c|c}
        \hline
        \hline
        Para.                                &$J/\psi\to\Sigma^+(p\pi^0)\bar\Sigma^-((\bar{p}\pi^0))$  &$J/\psi\to\Sigma^+(p\pi^0)\bar\Sigma^-((\bar{n}\pi^-))$    &$\psi(3686)\to\Sigma^+\bar\Sigma^-$     \\ \hline
        $\alpha$                           &$-0.508 \pm 0.006 \pm 0.004$  &$-0.5156 \pm 0.0030 \pm 0.0061$  &$0.682 \pm 0.030 \pm 0.011$\\      
        $\Delta\Phi$ (rad)              &$-0.270 \pm 0.012 \pm 0.009$   &$-0.2772 \pm 0.0044 \pm 0.0041$  &$0.379 \pm 0.070 \pm 0.014$\\        
        $\alpha_{0}$                      &$-0.998 \pm 0.037 \pm 0.009$  &--  &$-0.998 \pm 0.037 \pm 0.009$\\
        $\bar\alpha_{0}$                &$0.990 \pm 0.037 \pm 0.011$     &--  &$0.990 \pm 0.037 \pm 0.011$\\
        $\alpha_{+}$                      &--&$0.0481 \pm 0.0031 \pm 0.0019$    &--\\
        $\bar\alpha_{-}$                &--&$-0.0565 \pm 0.0047 \pm 0.0022$   &--\\
        $A_{CP}$                           &$-0.004 \pm 0.037 \pm 0.010$ &$-0.080 \pm 0.052 \pm 0.028$&--\\
        $R^{\Psi}$                          &$2.264 \pm 0.018 \pm 0.012$        &$2.287 \pm 0.009 \pm 0.019$ &$0.669 \pm 0.038 \pm 0.014$ \\ \hline \hline
        \end{tabular}}
        \label{table:ssb}
\end{table}
\subsubsection{$e^+e^-\to\Sigma^+\bar\Sigma^-$ near threshold}
With a total of 66 $pb^{-1}$ data samples taken at $\sqrt{s} =$ 2.396, 2.65 and 2.9 GeV, the measurements of the EMFF and
transverse polarization for $\Sigma^+$ hyperon in $e^+e^-\to\Sigma^+\bar\Sigma^-$ were first explored in a wide momentum transfer range at $q^2$=5.7-8.4 GeV~\cite{BESIII:2023ynq}. 
The relative phases between the electric and magnetic form factors are measured to be $\Delta\Phi < 0$ at $\sqrt{s} =$ 2.396 GeV,  while $\Delta\Phi >0$ at  $\sqrt{s} =$ 2.64  and 2.9 GeV, which implies that there may be at least one  $\Delta\Phi = 0$ existence between these energy points. This evolution would provide an important input for understanding the asymptotic behavior~\cite{Mangoni:2021qmd, Wang:2022zyc} and the dynamics of baryons.

\subsection{Hyperon physics in $\Xi$ decay}
\subsubsection{$J/\psi$, $\psi(3686)\to\Xi^-\bar\Xi^+$}
Using 1.3 billion $J/\psi$ and 448 million $\psi(3686)$ events taken by BESIII detector,  the probing for $CP$ symmetry and weak phases with entangled double-strange baryon pairs were presented in $J/\psi$, $\psi(3686)\to\Xi^-\bar\Xi^+$~\cite{BESIII:2021ypr, BESIII:2022lsz}.
Since the production and multi-step decays can be described by nine  kinematic variables and eight parameters related to the hyperon polarization and decay parameter in the helicity amplitude of $e^+e^-\to J/\psi$, $\psi(3686)\to\Xi^-\bar\Xi^+$ with $\Xi^-\to\pi^-\Lambda$ and $\Lambda\to p\pi^-$,  a nine dimension
 angular distribution analysis are implemented. Clear $\Xi^-$ spin polarizations are observed in $J/\psi$ and $\psi(3686)$ decays with the smallest significance of 7$\sigma$ as shown in Fig.~\ref{scatter_plot::xxb}. As for other parameters related to angular distribution, hyperon weak decay, 
strong and weak phase difference, they are measured to be consistent with the previous results, and 
with higher precision than the previous results, or are reported for the first time. Based on the measured decay parameters, 
three independent $CP$ observables in $\Xi^-$ decay are determined with an order of $10^{-3}$ , which is in good agreement with and compatible in precision to the most precise previous measurement in $\Lambda$ decay~\cite{BESIII:2018cnd}. Table~\ref{table:xxbar} summarizes the measured results.
This study provides a hunting ground for physics beyond the SM in strange baryon that is complementary to the results in kaon decay~\cite{Tandean:2003fr}.

Besides, compared with the measurements between $J/\psi$ and $\psi(3686)$ decays, both $\alpha_{\psi}$ and $\Delta\Phi$ are very different from each other, while other parameters and CPV observables are consistent with each other. It is indicated that the production mechanism of the entangled $\Xi^-\bar\Xi^+$ pair in both $J/\psi$ and $\psi(3686)$ decays are different and complicated.

\subsubsection{$J/\psi$, $\psi(3686)\to\Xi^0\bar\Xi^0$}
With the same strategy as used in $J/\psi$, $\psi(3686)\to\Xi^-\bar\Xi^+$~~\cite{BESIII:2021ypr, BESIII:2022lsz}, 
the spin polarization and $CPV$ study were also studied in the process $e^+e^-\to J/\psi$, $\psi(3686)\to\Xi^0\bar\Xi^0$ with $\Xi^0\to\pi^0\Lambda$ and $\Lambda\to p\pi^-$ based on 10 billion $J/\psi$ and 448 million $\psi(3686)$ events~\cite{BESIII:2023lkg, BESIII:2023drj}. 
Clear $\Xi^0$ spin polarization is observed in $J/\psi$ decay, while not in $\psi(3686)$ decay due to the statistics limitation as shown in Fig.~\ref{scatter_plot::xxb}.
The $\Xi^0$  and $\bar\Xi^0$ weak decay parameters are simultaneously measured with most accurate precision for the first time.
As for other parameters related to angular distribution parameters, 
strong and weak phase difference, and $CP$ observables, they are measured to be consistent and 
with higher precision than the previous results. Table~\ref{table:xxbar} summarizes the measured results. The $CP$ is  found to be still equal to zero within the uncertainty of 1$\sigma$.
By comparing the measurements between $J/\psi$ and $\psi(3686)$ decays, the results are consistent with each other within the uncertainty of 1$\sigma$.
\begin{figure}[!htbp]
\begin{center}
\includegraphics[width=0.42\textwidth]{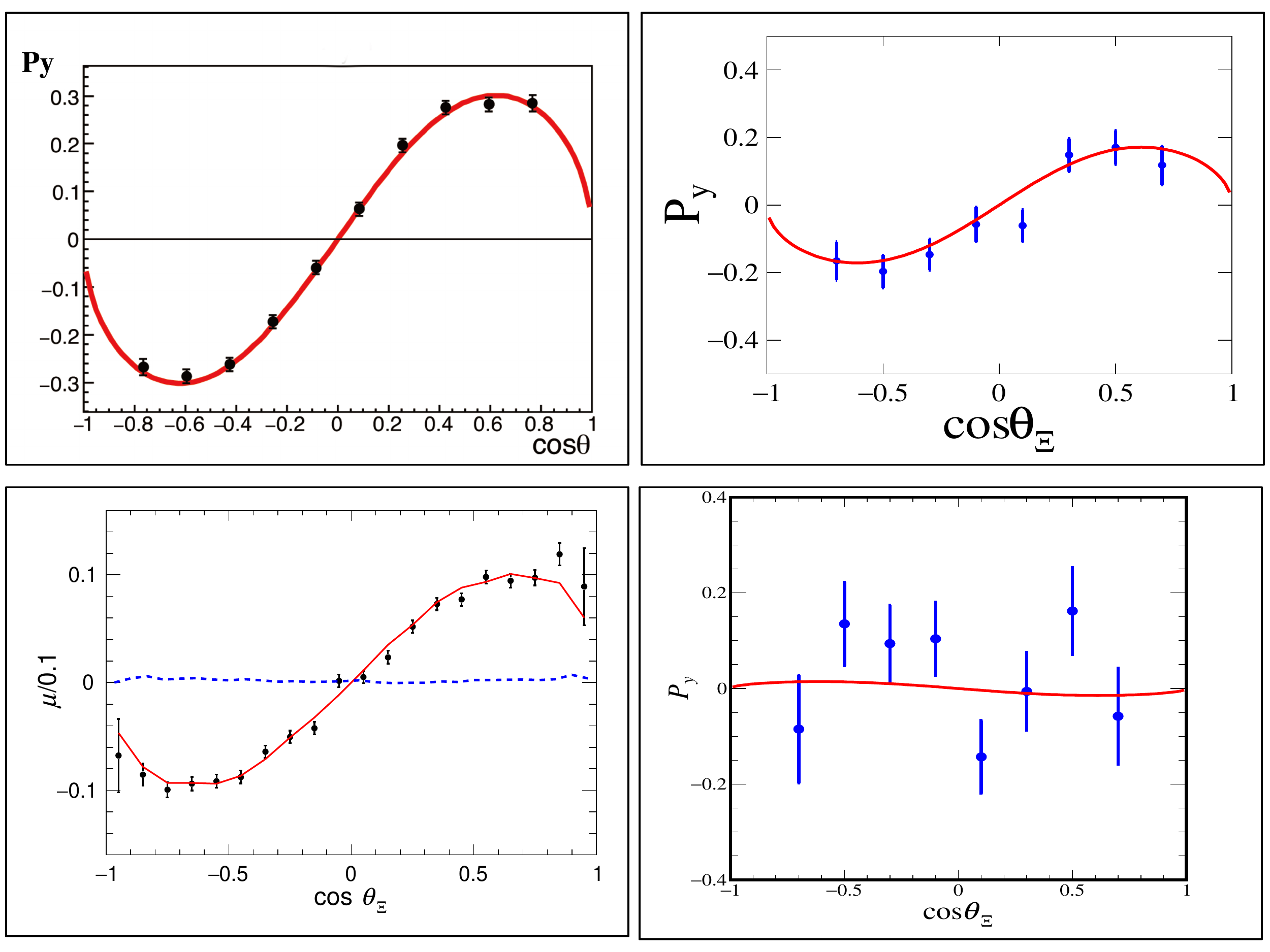}
\end{center}
\caption{\label{scatter_plot::xxb}The polarization ($P_y$) and moment $\mu(\cos\theta_{\Xi})$ as a function of $\cos\theta_{\Xi}$ in $J/\psi$, $\psi(3686)\to\Xi^-\bar\Xi^+$ (Top Left and Right) and $J/\psi$, $\psi(3686)\to\Xi^0\bar\Xi^0$ (Bottom Left and Right) . }
\label{scatter_plot::xxb}
\end{figure}

\begin{table*}[htbp]
        \centering
        \caption{Experiment parameters for angular distribution ($\alpha$), spin polarization ($\Delta\Phi$), decay parameters ($\alpha_{\Xi/\bar\Xi}$, $\alpha_{\Lambda/\bar\Lambda}$, $\phi_{\Xi/\bar\Xi}$,   $<\alpha_{\Xi/\Lambda}>$, $<\phi_{\Xi}>$) and $CP$ observable ($A^{\Xi, \Lambda}_{CP}$, $\Delta\phi^{\Xi}_{CP}$) in $J/\psi$ and $\psi(3686)$ decays.}
 \scalebox{0.9}{
        \begin{tabular}{l|c|c|c|c}
        \hline
        \hline
        Para.                            &$J/\psi\to\Xi^-\bar\Xi^+$~\cite{BESIII:2021ypr}   &$\psi(3686)\to\Xi^-\bar\Xi^+$~\cite{BESIII:2022lsz}    &$J/\psi\to\Xi^0\bar\Xi^0$~\cite{BESIII:2023lkg}    &$\psi(3686)\to\Xi^0\bar\Xi^0$~\cite{BESIII:2023drj}  \\ 
        \hline
        $\alpha$                                &$0.586\pm 0.012 \pm0.010$  &$0.693 \pm 0.048 \pm 0.049$    &$0.514  \pm 0.006 \pm  0.015$ &$0.665 \pm 0.086 \pm 0.081$\\     
        $\Delta\Phi$ (rad)           &$1.213\pm 0.046 \pm 0.016$ &$0.667 \pm 0.111 \pm 0.058$    &$1.168  \pm 0.019 \pm  0.018$ &$-0.050 \pm 0.150 \pm 0.020$\\        
        $\alpha_{\Xi}$                        &$-0.376\pm0.007\pm0.003$     &$-0.344 \pm 0.025 \pm 0.007$  &$-0.3750 \pm 0.0034 \pm 0.0016$&$-0.358 \pm 0.042 \pm 0.013$ \\
        $\alpha_{\bar\Xi}$                  &$0.371\pm0.007\pm0.002$    &$0.355 \pm 0.025 \pm 0.002$    &$0.3790 \pm 0.0034 \pm 0.0021$&$0.363 \pm 0.042 \pm 0.013$\\
        $\phi_{\Xi}$ (rad)                    &$0.011\pm0.019\pm0.009$    &$0.023 \pm 0.074 \pm 0.003$     &$0.0051 \pm 0.0096 \pm 0.0018$ &$0.027 \pm 0.117 \pm 0.011$\\
        $\phi_{\bar\Xi}$ (rad)               &$-0.021\pm0.019\pm0.007$  &$-0.123 \pm 0.073 \pm 0.004$     & $-0.0053 \pm 0.0097 \pm 0.0019$ &$-0.185 \pm 0.116 \pm 0.017$\\
        $\alpha_{\Lambda}$                &$0.757 \pm 0.011 \pm 0.008$&--                                                &$0.7551 \pm 0.0052 \pm 0.0023$&--\\
        $\alpha_{\bar\Lambda}$          &$-0.763 \pm 0.011 \pm 0.007$                                                     &--&$-0.7448 \pm 0.0052 \pm 0.0017$ &---\\
        $\xi_{P}-\xi_{S}$   ($\times10^{-2}$ rad)    &$-1.2 \pm 3.4 \pm 0.8$                                                &--&$0.0  \pm 1.7  \pm 0.2$ &--\\
        $\delta_{P}-\delta_{S}$  ($\times10^{-2}$ rad)  &$-4.0 \pm 3.3 \pm 1.7$   &$-20.0 \pm 13.0 \pm 1.0$ &$-1.3  \pm 1.7  \pm 0.4$&--\\
        $A^{\Xi}_{CP}$ ($\times10^{-3}$)          &$- 6.0 \pm 13.4 \pm 5.6$       &$-15.0 \pm 51.0 \pm 10.0$  &$-5.4 \pm 6.5 \pm 3.1$ &$-7.0 \pm 82.0 \pm 25.0$\\
        $A^{\Lambda}_{CP}$  ($\times10^{-3}$)   &$-4.8 \pm13.7 \pm 2.9$                     &--&$6.9 \pm 5.8 \pm 1.8$&--\\
        $\Delta\phi^{\Xi}_{CP}$  ($\times10^{-3}$ rad)  &$-3.7 \pm 11.7 \pm 9.0$             &$-50.0 \pm 52.0 \pm 3.0$ &$-0.1 \pm 6.9 \pm 0.9$&$-79.0 \pm 82.0 \pm 10.0$\\
        \hline
        \hline
        \end{tabular}}
        \label{table:xxbar}
\end{table*}
\section{Summary}
BESIII has been operating successfully since 2008 and by far has collected a large number of data samples in the $\tau$-charm physics region to date.
Many studies for spin polarization and $CPV$ in hyperon sector achieved, such as 
observation of spin polarization, the most accurate test of $CPV$ in $\Lambda$, $\Sigma^+$, $\Xi^0$ and $\Xi^-$ and so on.  
The accuracy for the test of CPV in $\Xi$ hyperon decay may move to the required sensitivity of SM in the future when using 10 billion $J/\psi$ events.
However, the challenge of achieving the required CPV sensitivity ($\sim10^{-4}$) still exists at BESIII.
\section*{FUNDING}
This work was supported by the National Natural Science Foundation of China under Contract Nos. 12075107, 12247101; 
and by the National Key Research and Development Program of China under Contract No. 2020YFA0406403; 
and by the 111 Project under Grant No. B20063.





\begin{thebibliography}{}
\bibitem{Sakharov:1967dj}
A. D. Sakharov,
\href{https://doi.org/10.1070/pu1991v034n05abeh002497} {Pisma Zh. Eksp. Teor. Fiz.\textbf{5}, 32-35 (1967)}.

\bibitem{Workman:2022ynf}
Particle Data Group,
\href{https://pdglive.lbl.gov/Viewer.action}{PTEP \textbf{2022}, 083C01 (2022)}. 

\bibitem{Peskin:2002mm}
M. E. Peskin,
\href{https://www.nature.com/articles/419024a}{Nature \textbf{419}, 24-27 (2002)}.

\bibitem{Cohen:1993nk}
A.G. Cohen, D. B. Kaplan and A. E. Nelson,
\href{https://www.annualreviews.org/doi/10.1146/annurev.ns.43.120193.000331}{Ann. Rev. Nucl. Part. Sci. \textbf{43}, 27-70 (1993)};
\href{https://arxiv.org/abs/hep-ph/9302210}{arXiv: hep-ph/9302210}.



\bibitem{BESIII:2012ghz}
BESIII Collaboration,
\href{https://doi.org/10.1103/PhysRevD.87.032007}{Phys. Rev. D \textbf{87}, 032007 (2013)}
\href{https://doi.org/10.1103/PhysRevD.87.059901}{[erratum: Phys. Rev. D \textbf{87}, 059901 (2013)]};
\href{https://arxiv.org/abs/1211.2283}{arXiv: 1211.2283}.



\bibitem{BESIII:2016ssr}
BESIII Collaboration,
\href{https://doi.org/10.1103/PhysRevD.93.072003}{Phys. Rev. D \textbf{93}, 072003 (2016)};
\href{https://arxiv.org/abs/1602.06754}{arXiv: 1602.06754}.


\bibitem{BESIII:2016nix}
BESIII Collaboration,
\href{http://dx.doi.org/10.1016/j.physletb.2017.04.048}{Phys. Lett. B \textbf{770}, 217-225 (2017)};
\href{https://arxiv.org/abs/1612.08664}{arXiv: 1612.08664}.




\bibitem{BESIII:2019dve}
BESIII Collaboration,
\href{https://doi.org/10.1103/PhysRevD.100.051101}{Phys. Rev. D \textbf{100}, 051101 (2019)};
\href{https://arxiv.org/abs/1907.13041}{arXiv: 1907.13041}.


\bibitem{BESIII:2019cuv}
BESIII Collaboration,
\href{https://doi.org/10.1103/PhysRevLett.124.032002}{Phys. Rev. Lett. \textbf{124}, 032002 (2020)};
\href{https://arxiv.org/abs/1910.04921}{arXiv: 1910.04921}.



\bibitem{BESIII:2020ktn}
BESIII Collaboration,
\href{https://doi.org/10.1103/PhysRevD.103.012005}{Phys. Rev. D \textbf{103}, 012005 (2021)};
\href{https://arxiv.org/abs/2010.08320}{arXiv: 2010.08320}.

\bibitem{BESIII:2021aer}
BESIII Collaboration,
\href{https://doi.org/10.1016/j.physletb.2021.136557}{Phys. Lett. B \textbf{820}, 136557 (2021)};
\href{https://arxiv.org/abs/2105.14657}{arXiv: 2105.14657}.


\bibitem{BESIII:2021gca}
BESIII Collaboration,
\href{https://doi.org/10.1103/PhysRevD.104.092012}{Phys. Rev. D \textbf{104}, 092012 (2021)};
\href{https://arxiv.org/abs/2109.06621}{arXiv: 2109.06621}.

\bibitem{BESIII:2021ccp}
BESIII Collaboration,
\href{https://doi.org/10.1103/PhysRevD.104.L091104}{Phys. Rev. D \textbf{104}, L091104 (2021)};
\href{https://arxiv.org/abs/2108.02410}{arXiv: 2108.02410}.

\bibitem{BESIII:2022mfx}
BESIII Collaboration,
\href{https://link.springer.com/article/10.1007/JHEP06(2022)074}{JHEP \textbf{06}, 74 (2022)};
\href{https://arxiv.org/abs/2202.08058}{arXiv: 2202.08058}.



\bibitem{BESIII:2022kzc}
BESIII Collaboration,
\href{https://doi.org/10.1103/PhysRevD.107.052003}{Phys. Rev. D \textbf{107}, 052003 (2023)};
\href{https://arxiv.org/abs/2212.03693}{arXiv: 2212.03693}.

\bibitem{BESIII:2023rse}
BESIII Collaboration,
\href{https://link.springer.com/article/10.1007/JHEP11(2023)228}{JHEP \textbf{11}, 228 (2023)};
\href{https://arxiv.org/abs/2309.04215}{arXiv: 2309.04215}.



\bibitem{Lee:1957qs}
T. D. Lee and C.~N.~Yang,
 \href{https://doi.org/10.1103/PhysRev.108.1645}{Phys. Rev. \textbf{108}, 1645-1647 (1957)}.

\bibitem{Liu:2023xhg}
H. Liu, J. Zhang and X.~Wang,
\href{https://doi.org/10.3390/sym15010214}{Symmetry \textbf{15}, 214 (2023)}.

\bibitem{BESIII:2018cnd}
BESIII Collaboration,
\href{https://doi.org/10.1038/s41567-019-0494-8}{Nature Phys. \textbf{15}, 631-634 (2019)};
\href{https://arxiv.org/abs/1808.08917}{arXiv: 1808.08917}.

\bibitem{Overseth:1969bxc}Olsen:1970vb
O. E. Overseth and S. Pakvasa,
\href{https://doi.org/10.1103/PhysRev.184.1663}{Phys. Rev. \textbf{184}, 1663-1667 (1969)}.

\bibitem{Olsen:1970vb}
S. Olsen, L. Pondrom, R. Handler, P. Limon, J. A. Smith and O. E. Overseth,
\href{https://doi.org/10.1103/PhysRevLett.24.843}{Phys. Rev. Lett. \textbf{24}, 843-847 (1970)}.

\bibitem{BESIII:2022qax}
BESIII Collaboration,
\href{https://doi.org/10.1103/PhysRevLett.129.131801}{Phys. Rev. Lett. \textbf{129}, 131801 (2022)};
\href{https://arxiv.org/abs/2204.11058}{arXiv: 2204.11058}.

\bibitem{BESIII:2023euh}
BESIII Collaboration,
\href{https://doi.org/10.1007/JHEP10(2023)081}{JHEP \textbf{10}, 081 (2023)};
\href{https://arxiv.org/abs/2303.00271}{arXiv: 2303.00271}.

\bibitem{BESIII:2021cvv}
BESIII Collaboration,
\href{https://doi.org/10.1103/PhysRevD.105.L011101}{Phys. Rev. D \textbf{105}, L011101 (2022)};
\href{https://arxiv.org/abs/2111.11742}{arXiv: 2111.1174}.

\bibitem{BESIII:2019nep}
BESIII Collaboration,
\href{https://doi.org/10.1103/PhysRevLett.123.122003}{Phys. Rev. Lett. \textbf{123}, 122003 (2019)};
\href{https://arxiv.org/abs/1903.09421}{arXiv: 1903.09421}.

\bibitem{BESIII:2020fqg}
BESIII Collaboration,
\href{https://doi.org/10.1103/PhysRevLett.125.052004}{Phys. Rev. Lett. \textbf{125}, 052004 (2020)};
\href{https://arxiv.org/abs/2004.07701}{arXiv: 2004.07701}.

\bibitem{BESIII:2023sgt}
BESIII Collaboration,
\href{https://doi.org/10.1103/PhysRevLett.131.191802}{Phys. Rev. Lett. \textbf{131}, 191802 (2023)};
\href{https://arxiv.org/abs/2304.14655}{arXiv: 2304.14655}.

 \bibitem{BESIII:2023ynq}
BESIII Collaboration,
\href{https://arxiv.org/abs/2307.15894}{arXiv: 2307.15894}.

 \bibitem{Mangoni:2021qmd}
A.~Mangoni, S.~Pacetti and E.~Tomasi-Gustafsson,
\href{https://doi.org/10.1103/PhysRevD.104.116016}{Phys. Rev. D \textbf{104}, 116016 (2021)};
\href{https://arxiv.org/abs/2109.03759}{arXiv: 2109.03759}.

\bibitem{Wang:2022zyc}
X. Wang and G. Huang,
\href{https://doi.org/10.3390/sym14010065}{Symmetry \textbf{14}, 65 (2022)}.

 \bibitem{BESIII:2021ypr}
BESIII Collaboration,
\href{https://doi.org/10.1038/s41586-022-04624-1}{Nature \textbf{606}, 64-69 (2022)};
\href{https://arxiv.org/abs/2105.11155}{arXiv: 2105.11155}.

\bibitem{BESIII:2022lsz}
BESIII Collaboration,
\href{https://doi.org/10.1103/PhysRevD.106.L091101}{Phys. Rev. D \textbf{106}, L091101 (2022)};
\href{https://arxiv.org/abs/2206.10900}{arXiv: 2206.10900}.

\bibitem{Tandean:2003fr}
J.~Tandean,
\href{https://doi.org/10.1103/PhysRevD.69.076008}{Phys. Rev. D \textbf{69}, 076008 (2004)};
\href{https://arxiv.org/abs/hep-ph/0311036}{arXiv: hep-ph/0311036}.

\bibitem{BESIII:2023lkg}
BESIII Collaboration,
\href{https://doi.org/10.1103/PhysRevD.108.L011101}{Phys. Rev. D \textbf{108}, L011101 (2023)};
\href{https://arxiv.org/abs/2302.09767}{arXiv: 2302.09767}.

\bibitem{BESIII:2023drj}
BESIII Collaboration,
\href{https://doi.org/10.1103/PhysRevD.108.L031106}{Phys. Rev. D \textbf{108}, L031106 (2023)};
\href{https://arxiv.org/abs/2305.09218}{arXiv: 2305.09218}.


\end{thebibliography}
\end{document}